\documentclass[12pt]{article}
\usepackage{epsfig}
\newskip\humongous \humongous=0pt plus 1000pt minus 1000pt

\newif\ifdtup

\def\etal{\hbox{\it et al.}} 

\def\VEV#1{\left\langle #1\right\rangle}

\def\pr#1{#1^\prime}

\def\beq{\begin{equation}}
\def\eeq{\end{equation}}

\def\beqn{\begin{eqnarray}}
\def\eeqn{\end{eqnarray}}
\relax

\catcode`\@=11

\@addtoreset{equation}{section}
\def\theequation{\thesection.\arabic{equation}}

\def\@normalsize{\@setsize\normalsize{15pt}\xiipt\@xiipt
\abovedisplayskip 14pt plus3pt minus3pt%
\belowdisplayskip \abovedisplayskip
\abovedisplayshortskip \z@ plus3pt%
\belowdisplayshortskip 7pt plus3.5pt minus0pt}

\def\small{\@setsize\small{13.6pt}\xipt\@xipt
\abovedisplayskip 13pt plus3pt minus3pt%
\belowdisplayskip \abovedisplayskip
\abovedisplayshortskip \z@ plus3pt%
\belowdisplayshortskip 7pt plus3.5pt minus0pt
\def\@listi{\parsep 4.5pt plus 2pt minus 1pt
     \itemsep \parsep
     \topsep 9pt plus 3pt minus 3pt}}

\@twosidetrue

\relax

\catcode`@=12

\catcode`\@=11

\def\section{\@startsection{section}{1}{\z@}{3.5ex plus 1ex minus
   .2ex}{2.3ex plus .2ex}{\large\bf}}

\def\thesection{\arabic{section}}

\def\appendix{\setcounter{section}{0}
 \def\thesection{APPENDIX \Alph{section}:}
 \def\theequation{\Alph{section}.\arabic{equation}}}

\def\ps@headings{\def\@oddfoot{}\def\@evenfoot{}
\def\@oddhead{\hbox{}\hfill
 \makebox[.5\textwidth]{\raggedright\ignorespaces --\thepage{}--
 \hfill {}}}  
\def\@evenhead{\@oddhead}
\def\subsectionmark##1{\markboth{##1}{}}
}

\ps@headings

\catcode`\@=12

\def\figcap{\section*{Figure Captions\markboth
 {FIGURECAPTIONS}{FIGURECAPTIONS}}\list
 {Fig. \arabic{enumi}:\hfill}{\settowidth\labelwidth{Fig. 999:}
 \leftmargin\labelwidth
 \advance\leftmargin\labelsep\usecounter{enumi}}}
 \relax
\def\tablecap{\section*{Table Captions\markboth
 {TABLECAPTIONS}{TABLECAPTIONS}}\list
 {Table \arabic{enumi}:\hfill}{\settowidth\labelwidth{Table 999:}
 \leftmargin\labelwidth
 \advance\leftmargin\labelsep\usecounter{enumi}}}
 \relax
\def\reflist{\section*{References\markboth
 {REFLIST}{REFLIST}}\list
 {[\arabic{enumi}]\hfill}{\settowidth\labelwidth{[999]}
 \leftmargin\labelwidth
 \advance\leftmargin\labelsep\usecounter{enumi}}}
 \relax

\catcode`\@=11

\relax

\relax
\def\pl#1#2#3{{\it Phys. Lett. }{\bf #1}(19#2)#3}

\def\prl#1#2#3{{\it Phys. Rev. Lett. }{\bf #1}(19#2)#3}

\def\prep#1#2#3{{\it Phys. Rep. }{\bf #1}(19#2)#3}
\def\pr#1#2#3{{\it Phys. Rev. }{\bf #1}(19#2)#3}

\relax

\catcode `@ 11
\def\biblabel#1{\if@filesw\immediate
\write\@auxout{\string\bibcite{#1}{\the\value{\@listctr }}}\fi}
\catcode `@ 12

\newcommand{\ccaption}[2]{
  \begin{center}
    \parbox{0.85\textwidth}{
      \caption[#1]{\small\it {#2}}}
  \end{center}    }
\def    \be             {\begin{equation}}
\def    \ee             {\end{equation}}
\def    \ba             {\begin{eqnarray}}
\def    \ea             {\end{eqnarray}}

\def    \=              {\;=\;}
\def    \frac           #1#2{{#1 \over #2}}

\def \ep{\epsilon}
\def \as   {\ifmmode \alpha_s \else $\alpha_s$ \fi}

\def\b0{b_0}

\def \mt   {\ifmmode m_{\rm t} \else $m_{\rm t}$ \fi}

\def \to   {\mbox{$\rightarrow$}}

\begin{document}
\begin{titlepage}
\nopagebreak
{\flushright{
        \begin{minipage}{4cm}
        CERN-TH/96-178  \hfill \\
        hep-ph/9607347\hfill \\
        \end{minipage}        }

}
\vfill
\begin{center}
{\LARGE 
{ \bf \sc  Momentum correlations in $Z^0\to b\bar b$ \newline
                   and the measurement of $R^0_b$ }}
\vskip .5cm
{\bf Paolo NASON\footnote{On leave of absence from INFN, Milan, Italy},}
\\
\vskip 0.1cm
{CERN, TH Division, Geneva, Switzerland} \\
\vskip .5cm
{\bf Carlo OLEARI}
\\
\vskip .1cm
{Univ. di Milano and INFN, Milan, Italy}
\end{center}
\nopagebreak
\vfill
\begin{abstract}
  We study the correlations of $b\bar b$ pairs produced in inclusive
  $Z^0$ decays, and their
  possible influence upon the determination of $R_b$.
\end{abstract}
\vskip 1cm
CERN-TH/96-178 \hfill \\
July 1996 \hfill
\vfill
\end{titlepage}
\section{Introduction}
The measured value of $R^0_b=\Gamma_{b\bar b}/\Gamma_{\rm had}$
in $Z^0$ decays is drawing considerable
attention in the physics community. On the one hand, according to all four
LEP experiments, it deviates significantly from the Standard Model prediction,
a result that has remained quite stable in the last few years
\cite{DELPHI1}--\cite{DELPHI3}.
On the other hand, it is a considerably difficult measurement,
and it is certainly disturbing that the only deviation from the Standard Model
is found in a quantity that
is affected by complicated strong-interaction effects.
In the present paper we examine one possible source of systematic errors
that may affect the measurement of $R^0_b$, whose origin is purely
dynamical. In several experimental techniques for the measurement of $R^0_b$,
the tagging efficiency is extracted from data by comparing the
sample of events in which only one $b$ has been tagged, with the one in which
both $b$'s are observed. If the production characteristics of the $b$
and the $\bar b$ were completely uncorrelated, this method would yield
the exact answer with no need for corrections. Of course, other correlations
of an experimental nature should be properly accounted for,
but their discussion
is outside the scope of the present theoretical paper.
Standard QCD gluon emission generates a correlation of the
quark--antiquark momenta of the order of $\as$.
Other dynamical effects, such as the production of heavy quark pairs
via a gluon-splitting mechanism, may affect the measurement.
However, they are to some extent better
understood. In the first place they tend to give soft
heavy quarks, and they are therefore easily eliminated.
Furthermore, experimental studies of these production mechanisms
have begun to appear~\cite{GluonSplitting}.

The aim of the present work is to study the dynamical correlation
of the heavy flavoured quark--antiquark pair, with particular emphasis
upon its impact on the determination of $R_b$. We will use
leading-order QCD formulae throughout, emphasizing the cases in which
a higher-order calculation would be desirable and possible.
We found that a particularly relevant quantity is the average
momentum correlation $r$ defined as
\beq\label{defr}
r=\frac{\VEV{x_1 x_2}-\VEV{x}^2}{\VEV{x}^2}\,,
\eeq
where $x_{1(2)}$ are the Feynman $x$ of the produced $B(\bar{B})$ mesons,
$x_{1(2)}=p_{B(\bar{B})}/p^{(\rm max)}_B$, and
$\VEV{x}=\langle x_{1(2)} \rangle$.
In fact, we will show that under the assumption that the detection
efficiency for a $B$ meson
is nearly proportional to its momentum, the quantity $r$ represents
the relative correction to $R_b$ due to momentum correlations.
The quantity $r$ has a well-behaved perturbative expansion, starting
at order $\as$, in spite of the fact that the perturbative
expansions of $\VEV{x_1 x_2}$ and $\VEV{x}$ are not well-behaved,
since their coefficients are enhanced by logarithms of the total
annihilation energy over the heavy-quark mass. Our perturbative
calculation for $r$ gives a value of the order of 1\%.
Power-suppressed corrections may still
affect this quantity to some extent. If power corrections of the
order of $1/Q$ are present, they may turn out to be of the same
order of the perturbative value at LEP energy.
Monte Carlo models~\cite{MCmodels}
seem to support this possibility~\cite{MontecarloCalc}.

In general, the QCD formalism allows us to compute
the double differential cross section for the inclusive production
of the heavy-flavour pair, using as input the value of $\as$ and
the fragmentation function of heavy-flavoured mesons.
It is easy to compute such quantities in leading order.
The leading order result may be improved on one hand by including
next to leading corrections, and by resumming terms that are enhanced
logarithmically near the end of the phase
space (i.e. for $x_{1(2)}\to 1$). At present  these calculations
are not available.

Shower Monte Carlo models include to some extent perturbative effects
of leading order, and effects that are enhanced near the end of
the phase space. At the parton level, they should therefore provide
a description that is compatible with the leading-order calculations
reported in the present paper. Some recent studies \cite{MontecarloCalc}
support this conclusion. On the other hand, fixed order matrix elements
Monte Carlo programs should be used cautiously when computing correlations,
because of problems arising in the truncation of the perturbative expansion.

The rest of the paper is organized as follows. In Section~\ref{linep}
we discuss the average momentum correlation, and how it
affects the determination of $R_{b}$. In Section~\ref{dsigma}
we describe the leading-order calculation of the double inclusive
cross section for heavy flavour production. Finally,
we give our conclusions in Section~\ref{Concl}.

\section{The average momentum correlation}\label{linep}
We begin by considering the simple case in which 
the efficiency for tagging a $B$ meson is a linear function
of its momentum.
This simplifying assumption allows us to make very precise statements
about the correlation. Furthermore, it is not extremely far from
reality, in the sense that the experimental tagging efficiency is often
a growing function of the $B$ momentum.
Let us therefore assume
\beq\label{tagefs}
\ep_b=C x_1\,,\quad \ep_{\bar{b}}=C x_2\,,
\eeq
where $\ep_b$ is the tagging efficiency.
We assume that in our ideal detector the detection efficiency is not
influenced by the presence of another tag. Therefore the total
number of tags is then given by
\beq
\frac{N_1}{N}=R^0_b
C\left(\VEV{x_1}+\VEV{x_2}\right)=R^0_b \, 2C\VEV{x}\,,
\eeq
where $x_2$ is the Feynman $x$ of the produced $\bar B$ meson,
and with $\VEV{x}$ we indicate the common value of their average.
Observe that the in the above equation $N_1$ is the number of tags,
given by the number of events with one tag plus twice the number
of events with two tags.
On the other hand, the number of events with two tags will be given by
\beq
\frac{N_2}{N}=R^0_b C^2\VEV{x_1 x_2}
=R^0_b\,\left(C\VEV{x}\right)^2\times(1+r),
\eeq
where $r$ is defined in eq.~(\ref{defr}).
We can then extract $R^0_b$
\beq
R^0_b=\frac{N_1^2}{4 N N_2}\times(1+r)\;.
\eeq
The quantity $r$ cannot be measured; one therefore has to compute
it in order to determine $R^0_b$.

\begin{figure}
\centerline{\epsfig{figure=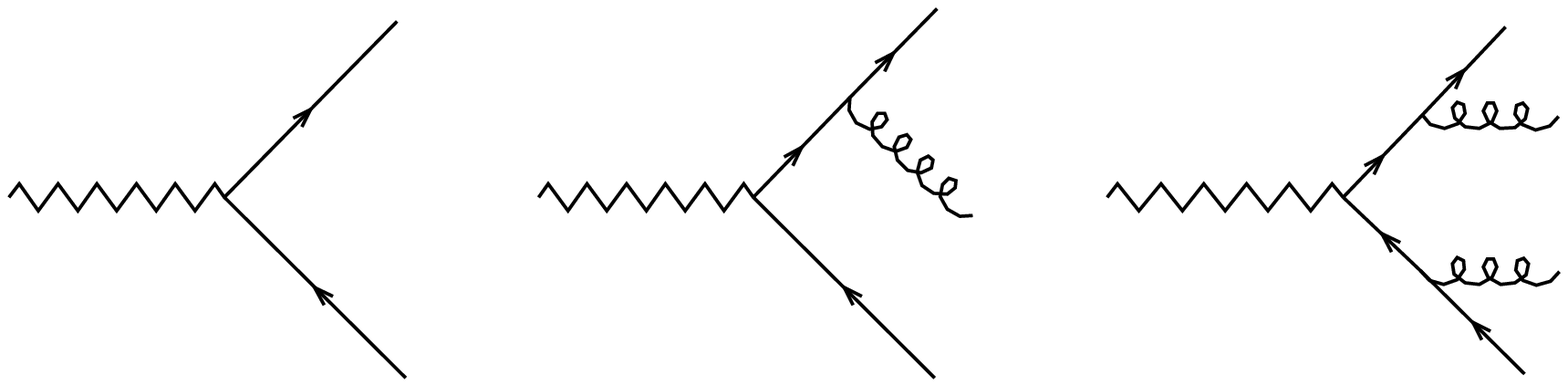,width=0.7\textwidth,clip=}}
\ccaption{}{ \label{ztobb}
Some of the diagrams contributing to the process $Z\to b\bar b$.
Virtual diagrams are not shown. }
\end{figure}

The basic QCD processes for the production
of a $b\bar b$ pair are depicted in fig.~\ref{ztobb}.
Let us focus on the process up to the order $\as$.
\newcommand\xe{{\bar x}}
The final state is fully defined in this case by the variables $\xe_1=2e_b/E$
and $\xe_2=2e_{\bar b}/E$, where $E$ is the total centre-of-mass energy.
The gluon momentum fraction $x_3$ is given by $2-\xe_1-\xe_2$.
We can write the differential cross section, normalized to 1, as
\beq
\frac{d \sigma}{d \xe_1 d \xe_2}=
\delta(1-\xe_1)\delta(1-\xe_2)+\as \frac{d \sigma^{(1)}}{d \xe_1 d \xe_2}\,,
\eeq
The term of order $\as$ is singular when $\xe_1$ and $\xe_2$ approach 1.
In fact, in this limit the emitted gluon is soft, and one expects a $1/x_3$
singularity. Furthermore, virtual corrections, concentrated
in the region $\xe_1=1$ and $\xe_2=1$ are also infrared-divergent.
Since our differential cross section is normalized to 1,
it is easy to write down a formula for the average of
some physical quantity $G(\xe_1,\xe_2)$ without having to explicitly
introduce the virtual corrections. We simply write
\beq\label{orderalfa}
\VEV{G(\xe_1,\xe_2)}=G(1,1)+\as \int d\xe_1\,d\xe_2\,
\frac{d \sigma^{(1)}}{d \xe_1 d \xe_2}\,\left(G(\xe_1,\xe_2)-G(1,1)\right)\;.
\eeq
The integral is extended in the appropriate Dalitz region.
The subtraction term under the integral sign embodies the
effects of the virtual corrections, and of the normalization
to the total cross section. Thus, if we choose $G(\xe_1,\xe_2)=1$ we
clearly get 1, which is our normalization.
Appropriate formulae for $d \sigma^{(1)}/d \xe_1 d \xe_2$ can be found
for example in ref.~\cite{Jersak}.
Formula (\ref{orderalfa}) can be easily implemented in a computer program.
We computed the following quantities
\beqn
c_1&=&\int d\xe_1\,d\xe_2\,
\frac{d \sigma^{(1)}}{d \xe_1 d \xe_2}\,\left(x_1-1\right)\;,
 \\
d_1&=&\int d\xe_1\,d\xe_2\,
\frac{d \sigma^{(1)}}{d \xe_1 d \xe_2}\,\left(x_1\,x_2-1\right)\;,
 \\
e_1&=&\int d\xe_1\,d\xe_2\,
\frac{d \sigma^{(1)}}{d \xe_1 d \xe_2}\,
\left(x_1\,x_2\,{\rm cut}(x_1,x_2)-1\right)\;,
\eeqn
where
\beq
{\rm cut}(x_1,x_2)=\theta\left({\rm max}(x_1,x_2)-x_3\right)\,.
\eeq
The theta function in the last equation corresponds to the requirement
that in doubly tagged events the $b$ and the $\bar b$ are in opposite
hemispheres with respect to the thrust axis. We have
\beq
\VEV{x}=1+\as c_1,\quad \VEV{x_1 x_2}=1+\as d_1,\quad
\VEV{x_1 x_2\,{\rm cut}(x_1,x_2)}=1+\as e_1.
\eeq
We find, for $m_b=5\;$GeV,
\beq
 c_1=-1.142,\quad d_1=-2.185,\quad e_1=-2.223\,.
\eeq
Observe that the coefficients we found are quite large, since one would
have naively expected them to be of the order of $1/\pi$. This is because
large logarithms $\log E/m_b$ arise in the coefficients, thus making the
perturbative expansion for the above quantities quite unreliable.
The quantities $r$ and $r^\prime$, however, have small perturbative
coefficients
\beqn
r&=&\frac{\VEV{x_1 x_2}-\VEV{x}^2}{\VEV{x}^2}=\as(d_1-2c_1)=0.099\,,\as
\nonumber \\ \label{corres}
r^\prime&=&\frac{\VEV{x_1 x_2\,{\rm cut}(x_1,x_2)}-
          \VEV{x}^2}{\VEV{x}^2}=\as(e_1-2c_1)=0.061\,\as\,.
\eeqn
Observe that the coefficients are even below the expected
magnitude of $1/\pi$.
A better illustration of the presence of large logarithms, and of their
cancellation in the quantities $r$ and $r^\prime$ is given in
table~\ref{tab:logdep}, where the coefficients $c_1$, $d_1$ and $e_1$
have been computed for $m=10$, 1, and 0.1~GeV. While $c_1$, $d_1$ and $e_1$
grow as the mass decreases, $r$ and $r^\prime$ approach a finite limit.
\begin{table}[htbp]
  \begin{center}
    \leavevmode
    \begin{tabular}{|c|c|c|c|c|c|}
      \hline
      $m$     & $c_1$  & $d_1$  & $e_1$  & $r$   & $r^\prime$ \\
      \hline
      10 GeV  & -0.773 & -1.460 & -1.486 & 0.086$\,\as$ & 0.060$\,\as$ \\
      \hline
      1 GeV   & -2.053 & -4.000 & -4.045 & 0.106$\,\as$ & 0.061$\,\as$ \\
      \hline
      0.1 GeV & -3.348 & -6.590 & -6.636 & 0.106$\,\as$ & 0.060$\,\as$ \\
      \hline
    \end{tabular}
    \caption{Mass depencence of the coefficients $c_1$, $d_1$ and $e_1$.}
    \label{tab:logdep}
  \end{center}
\end{table}
This is due to the fact that
the large logarithms present in $e_1$, $d_1$ and $e_1$
cancel in the linear combinations appearing in $r$. It is easy to prove
that this cancellation must occur to all orders in perturbation theory.
In fact, according to the factorization theorem, the
double inclusive cross section $b \bar b$ production, in the limit
of $E\gg m_b$, can be written as
\beq\label{doubleinc}
\frac{d\sigma}{dx_1\,dx_2}=\int \frac{d\hat\sigma}{dy_1\,dy_2}
\;D(z_1)\;D(z_2)\; \delta(y_1\,z_1-x_1)\;\delta(y_2\,z_2-x_2)\;
dy_1\,dz_1\,dy_2\,dz_2 \,,
\eeq
where (in the limit $E/m_b\to\infty$) $\hat\sigma$
has a perturbative expansion in $\as$ with
finite coefficients
(observe that the distinction between the momentum- or energy-defined
Feynman $x$ becomes irrelevant in the limit we are considering here).
The divergent terms are all
absorbed in the fragmentation functions $D(z)$. We then have
\beq\label{fullavx}
\VEV{x}=\int dx_1 dx_2\; x_1 \frac{d\sigma}{dx_1 dx_2}
=\left(\int dz\,z D(z)\right)
\int dy_1dy_2\; y_1 \frac{d\hat\sigma}{dy_1dy_2}\,,
\eeq
\beq
\VEV{x_1 x_2}=\int dx_1 dx_2\; x_1 x_2\; \frac{d\sigma}{dx_1 dx_2}
=\left(\int dz\,z D(z)\right)^2
\int dy_1 dy_2\; y_1 y_2\;\frac{d\hat\sigma}{dy_1 dy_2}\,.
\eeq
In the ratio $\VEV{x_1\,x_2}/\VEV{x}^2$ (and therefore in $r$)
the integral containing $D$ cancels.
Thus, the perturbative coefficients of $r$ are finite in the limit
$E/m_b\to \infty$. Observe that in the derivation of eq.~(\ref{fullavx})
we have assumed the relation
\beq
\int D(z)\,dz=1\,,
\eeq
which is appropriate when we can neglect the secondary production
of $b\bar b$ pairs via gluon splitting.

Since the cancellation of large logarithms takes place order by
order in perturbation theory, it would be wrong to include incomplete
higher-order corrections in the expression for $r$. Thus, for example,
if we write
\beq
r=\frac{1+d_1\as-(1+c_1\as)^2}{1+c_1\as}=(d_1-2c_1)\as+c_1(3c_1-2d_1)\as^2
+\cdots\,,
\eeq
the coefficient in front of the term of order $\as^2$ is very large
when compared to the coefficient in front of the term of order $\as$.
This is due to the fact that there are missing terms in the
${\cal O}(\as^2)$. If a complete ${\cal O}(\as^2)$ calculation is performed,
these large terms cancel.
Although a calculation of the next-to-leading term is in principle
possible, it is not yet available.
A calculation of the process of $b {\bar b} g g$
production has been given in ref.~\cite{Maina}, but the virtual corrections
for the $b {\bar b} g$ process in the case of massive quarks are not available.
In order to give an estimate of the next-to-leading effects,
and to verify the cancellation of the large terms in the ${\cal O}(\as^3)$
coefficients, we have computed the ${\cal O}(\as^2)$ contribution to $r$
coming from the real emission process $b {\bar b} g g$ alone.
We would like to emphasise that these results can be considered at most
as an estimate of the order of magnitude of the ${\cal O}(\as^2)$
coefficient, since the virtual terms are missing.
Appropriate subtraction terms have been included in order to
regulate the soft singularities. We get
\beq
c_2=-2.845\,,\quad d_2=-4.595\,,\quad e_2=-4.499\,.
\eeq
We have
\beqn
r&=&(d_1-2c_1)\as+(d_2-2c_1d_1-2c_2+3c_1^2)\as^2+{\cal O}(\as^3)
\nonumber \\
&=&0.099 \as+0.017\as^2+{\cal O}(\as^3)\,,
\nonumber \\
r^\prime&=&(e_1-2c_1)\as+(e_2-2c_1e_1-2c_2+3c_1^2)\as^2+{\cal O}(\as^3)
\nonumber \\
&=&0.061 \as+0.026\as^2+{\cal O}(\as^3)\,,
\eeqn
which shows a well-behaved perturbative expansion.
Notice that if we do not include the $c_2$, $d_2$ and $e_2$ terms,
the coefficients of the term of order $\as^2$ are an order of magnitude
larger than the coefficient of the term of order $\as$,
which again shows the importance of including all enhanced terms.

The correlation we computed is a small effect, of the order
of 1\%. In fact, we have to decide what is the scale at which $\as$
ought to be evaluated in eqs.~(\ref{corres}). In view of our discussion
on the cancellation of large logarithms, the correlation should
be dominated by large momenta, and therefore the appropriate scale
should be of the order of $E$. In fact, experience with jet physics
suggests the use of a somewhat smaller value. For example, the range
$0.12<\as<0.16$ gives a value of $r^\prime$ between 0.7 and 1~\%.

\section{Double inclusive cross section}\label{dsigma}
\begin{figure}
\centerline{\epsfig{figure=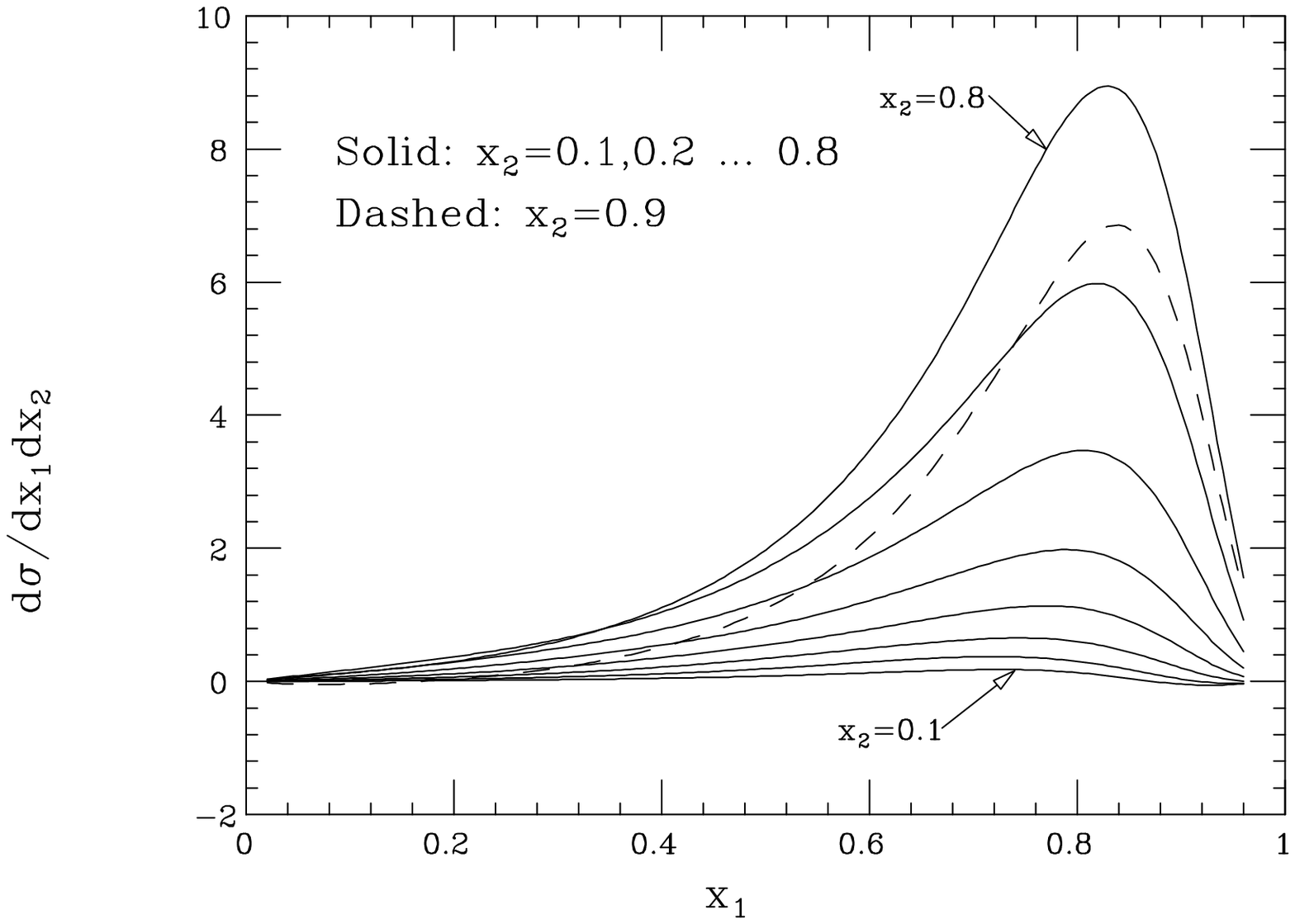,width=0.7\textwidth,clip=}}
\ccaption{}{ \label{sigdx1x2}
Double inclusive cross section $d \sigma/dx_1\,dx_2$, plotted as
a function of $x_1$ for several values of $x_2$. }
\end{figure}

Up to now, we have assumed in our discussion that the efficiency
is linear in the $B$ momentum.
Even in the more realistic case, in which the efficiency is
a more complicated function of the kinematic variables,
it is possible to compute the inclusive cross section for
the production of a $b\bar b$ pair, provided one also knows the
$B$ fragmentation function, which is to some extent
measured at LEP. The appropriate formula is given in eq.~(\ref{doubleinc}).
The expression for the short-distance cross section $\hat\sigma$
up to the order $\as$ is
\beq\label{doubleinchat}
\frac{d\hat\sigma}{dy_1\,dy_2}=\delta(1-y_1)\delta(1-y_2)
+\frac{2\as}{3\pi}\left.\frac{y_1^2+y_2^2}{(1-y_1)(1-y_2)}\right|_+\;.
\eeq
The $+$ distribution sign specifies the way that the singularities
at $y_1=1$ and $y_2=1$ should be treated. For any smooth function
of $y_1,\,y_2$ we define
\beqn \label{doubleplus}
&&\int_{y_1+y_2>1}
 dy_1\,dy_2\, \left.\frac{y_1^2+y_2^2}{(1-y_1)(1-y_2)}\right|_+
G(y_1,y_2)=
\\ \nonumber
&&\int_{y_1+y_2>1} dy_1\,dy_2\, \frac{y_1^2+y_2^2}{(1-y_1)(1-y_2)}
\left(G(y_1,y_2)-G(1,y_2)-G(y_1,1)+G(1,1)\right)\;.
\eeqn
Eqs. (\ref{doubleinc}) and (\ref{doubleinchat}) consistently gives
\beq\label{schemecond}
\frac{d\sigma}{dx_1}=\int dx_2 \frac{d\sigma}{dx_1\,dx_2}=D(x_1)\,.
\eeq
The above equation fixes the factorization scheme to be the
{\it annihilation scheme} as defined in ref.~\cite{NasonWebber}.
Observe that, in this case, the choice of scheme, i.e. eq.~(\ref{schemecond}),
fixed unambiguously the result without the need of computing
explicitly the virtual corrections. In fact, the most general formula
for the short distance cross section is obtain by adding to
eq.~(\ref{doubleinchat}) terms of the form
\beq
\frac{\as}{\pi}\left[\delta(1-y_1) f(y_2)+f(y_1)\delta(1-y_2)
 + g\delta(1-y_1)\delta(1-y_2)\right]\,,
\eeq
where $f$ is a generic function (in general a distribution) of one variable
and g is a number. However,
in order for eq.~(\ref{schemecond}) to be respected, these terms
must be absent.
From eqs. (\ref{doubleinc}), (\ref{doubleinchat}) and (\ref{doubleplus})
we immediately derive the ${\cal O}(\as)$ formula
\beqn &&
\frac{d\sigma}{dx_1 d x_2}\;=\;
D(x_1)D(x_2)+\frac{2\as}{3\pi}
\int_0^1 dy_1 dy_2
\;\theta(y_1+y_2-1)\;\frac{y_1^2+y_2^2}{(1-y_1)(1-y_2)}\;\times
\nonumber \\ &&
\Bigg[ D\left(\frac{x_1}{y_1}\right)D\left(\frac{x_2}{y_2}\right)
\frac{1}{y_1 y_2}
\,\theta(y_1-x_1)\,\theta(y_2-x_2)
-D(x_1)D\left(\frac{x_2}{y_2}\right)\frac{1}{y_2}\,\theta(y_2-x_2)
\nonumber \\ &&
-D\left(\frac{x_1}{y_1}\right)D(x_2)\frac{1}{y_1}\,\theta(y_1-x_1)
+D(x_1)D(x_2)\Bigg]\,.
\eeqn
As an illustration, we plot in fig.~\ref{sigdx1x2}
the double inclusive cross section
$d \sigma/dx_1\,dx_2$ as
a function of $x_1$ for several values of $x_2$.
We use the Peterson parametrization of the fragmentation function
\beq
D(z)=N_\ep\frac{z(1-z)^2}{((1-z)^2+z\ep)^2},
\eeq
where $N_\ep$ is fixed by the condition $\int D(x) dx=1$.
We used the value $\ep=0.04$, which gives $\VEV{x}=0.70$,
and $\as=0.12$.
The positive momentum correlation is quite visible in the figure.
As $x_2$ increases, the peak of the distribution in $x_1$ also
moves towards larger values.
Using the above formula, we can again compute $r$. The result
should not depend upon the choice of the fragmentation
function. In this case, the quantity $\VEV{x}$ does not receive
corrections of order $\as$. We get
\beq
\VEV{x}=0.7036\,,\quad\VEV{x_1 x_2}=0.4950+0.0525\,\as\,,\quad
r=0.1061\as\,,
\eeq
which is close to the value previously obtained $r=0.099$.
In fact, the difference is due to mass effects, as can be seen
from table~\ref{tab:logdep}.

There are several ways in which the above leading order calculation
could be improved in principle.
First of all, in the region of $x_1,\,x_2$ near 1, large logarithms
of $1-x$ arise, and they could be resummed to all order in perturbation
theory. It would also be desirable to include next-to-leading
corrections to the partonic cross section. At this moment,
these corrections have not yet been computed.

\section{Conclusion}\label{Concl}
Monte Carlo programs implement to some extent some of the QCD
dynamics discussed in the present work.
We therefore expect that, in general, at the parton level, they
should give results consistent with the QCD calculation
reported here, at least for quantities like $r$, in which large
logarithmic effects cancel. Observe that the computation of $r$
must be performed either consistently at some fixed order,
or including all logarithmically enhanced terms.
Failure to do so may expose large, uncancelled logarithmic terms,
and it may therefore lead to a wrong result.
Shower Monte Carlo programs include all logarithmically enhanced terms.
On the other hand,
matrix element Monte Carlo programs use a truncated perturbative
expansion. They can be used to compute $r$
only if each fixed-order contribution can be isolated,
and all terms
of order higher than the accuracy of the Monte Carlo can be thrown away.

The QCD factorization theorem guarantees that in quantities like $r$
large logarithmic effects, as well as hadronization effects,
should cancel. This statement is valid in a leading-twist sense.
Thus, it is possible that power-suppressed corrections to $r$ are
present. A correction of the order of $1/Q$ would be
a 1\% effect, comparable to the ${\cal O}(\as)$ corrections.
It is difficult to say anything about the importance of power corrections,
since our knowledge of the hadronization mechanism is quite poor.
One has therefore to rely upon hadronization models, which are commonly
implemented in Monte Carlo programs. The formalism presented in this paper
can also be applied to the computation of other processes, such as,
for example, the correlation in the inclusive production of strange mesons
in the large-$x$ region. These processes may provide
a testing ground for the theoretical predictions,
and may help us gain confidence on the reliability
of the hadronization models implemented in the Monte Carlo programs.
\vskip 0.5 cm
\noindent
{\bf Acknowledgements}\newline\noindent
We wish to thank G.~Altarelli, A.~Blondel, G.~Cowen, L.~L\"onnblad,
M.~Mangano, C.~Mariotti, K.~M\"onig, M.~Seymour, J.~Steinberger and B. Webber
for useful discussions.

\end{document}